# Toward Human-Centered Human–AI Interaction: Advances in Theoretical Frameworks and Practice

Zaifeng Gao[1], Yuanxiu Zhao[1], Hanxi Pan[1], Wei Xu[1,2]*

1. Department of Psychology and Behavioral Sciences, Zhejiang University, Hangzhou 310058, China
2. Center for Psychological Sciences, Zhejiang University, Hangzhou 310058, China

**Abstract**

With the rapid development of artificial intelligence (AI), machines are increasingly evolving into intelligent agents capable of autonomous decision-making, and the human–machine relationship is shifting from traditional "human–computer interaction" toward a new paradigm of "human–AI collaboration." However, technology-centered approaches to AI development have gradually revealed limitations such as fragility, bias, and low explainability, highlighting the urgent need for human-centered AI (HCAI) design philosophy. As a systems engineering approach, the successful implementation of HCAI depends critically on the design and optimization of high-quality human–AI interaction (HAII). This paper systematically reviews our research team's nearly decade-long exploration and practice in the field of HCAI. At the level of research vision, we were among the first in China to systematically propose HAII as an emerging interdisciplinary field and to develop a human-centered conceptual framework for human–AI collaboration. At the theoretical level, we introduced frameworks for human–AI joint cognitive systems, models of team-level situation awareness among intelligent agents, and shared social understanding models, forming a relatively comprehensive theoretical system. At the methodological level, we established a hierarchical HCAI framework and a taxonomy of five major categories of implementation methods. At the application level, we conducted a series of studies in domains such as autonomous driving, intelligent aircraft cockpits, and trust in human–AI collaboration, empirically validating the effectiveness of the proposed frameworks. Looking ahead, research on HCAI and HAII must continue to advance along three dimensions: theoretical deepening, methodological innovation, and application expansion, promoting the development of an intelligent society that is human-centered and characterized by harmonious human–AI coexistence.

**Keywords**: human-centered; human–AI interaction; human–AI collaboration; joint cognitive systems

# 向以人为中心的人-AI 交互：理论体系与实践进展

高在峰[1] 赵苑秀[1] 潘晗希[1] 许为[1,2]*

1. 浙江大学心理与行为科学系，杭州 310058
2. 浙江大学心理科学研究中心，杭州 310058

**摘要**

随着人工智能（Artificial Intelligence, AI）的快速发展，机器正逐步演化为具备自主决策能力的智能体，人机关系由传统的"人机交互"迈向"人-AI 协作"的新模式。然而，以技术为中心的 AI 开发模式逐渐暴露出脆弱性、偏见和可解释性低等局限，凸显了以人为中心的 AI（Human-Centered AI, HCAI）设计理念的迫切性。HCAI 作为系统工程理念，其成功落地依赖于高质量的人-AI 交互（Human-AI interaction, HAII）的设计与优化。本文系统梳理了我们团队在 HCAI 领域近十年的探索与实践。在研究理念上，我们在国内系统提出 HAII 这一跨学科新兴领域，并构建了以人为核心的人智协作概念框架；在基础理论上，提出了人智协同认知系统框架、智能体团队态势感知模型和共享社会理解模型，形成较为完整的理论体系；在方法论上，构建了层级式 HCAI 框架及五大类实现方法体系；在应用实践上，围绕自动驾驶、智能飞机驾驶舱和人智协作信任等场景场景开展系列研究，检验了框架有效性。展望未来，HCAI 和 HAII 研究仍需在理论深化、方法创新和应用拓展三方面持续发力，以推动构建以人为中心、人机和谐共生的智能社会。

## 引言

第三次 AI 浪潮正推动机器从被动工具跃升为具备自主决策的智能伙伴[1–4]，人机系统由此进入“协同认知”新范式——人与 AI 作为认知队友，以共享信任、态势与控制权实现双向动态交互[2,5]。这一趋势拓展了传统"人机交互"，形成"人机交互+人-AI 协作（简称人智协作）"的新型人机关系模式[2,5–9]，与任向实所提出的"人机共协计算"[10]相呼应，标志着智能时代人机关系的根本性跃迁。然而，“技术至上”的 AI 开发模式暴露出系统裂痕：AI 存在脆弱性、偏见、低解释性、因果模型缺失和伦理隐忧[11–16]。其中，"黑盒"效应尤为严重[17,18]，使用户难以理解和预测系统行为，甚至可能引发失控。2025 年 5 月，OpenAI 推理模型 o3 拒绝执行关机指令并修改系统代码[19]，提示当算法出于“自保”而违抗指令时，传统工具论瞬间失效。历史学家赫拉利警示，掌握语言密钥的 AI 或可撕裂社会共识[20]。这些风险的根源在于当前 AI 开发过度聚焦技术性能，忽视了人的需求、价值与认知特征。

因此，HCAI 已刻不容缓。HCAI 主张将人的需求、价值与能力置于核心，以确保决策符合人类期望，增强协作与信任，实现赋能而非替代[21]。许为等 26 位学者总结了 HCAI 面临的六大挑战：人类福祉、伦理责任、隐私保护、人本设计、治理监管和 HAII，为后续研究提供了方向[22]。HCAI 已发展为涵盖伦理、认知、社会科学和设计方法的系统工程理念，成为构建安全、可信、透明、增益公共善的智能生态的重要指导思想。作为整体性理念，HCAI 的有效落地高度依赖 HAII。HAII 既是 HCAI 的核心要素，也是其实现途径和跨学科合作平台。作为国内较早关注该领域的团队，我们围绕 HCAI 与 HAII 已开展近十年的探索与实践。本文在此基础上，从研究理念、基础理论、方法论和应用实践四个维度，总结相关进展并展望未来发展方向。

## 1 研究理念

本节旨在阐述 HCAI 理念的发展历程，并介绍我们团队提出的 HAII 学科定位和人类主导的协作框架，为后续理论研究和方法论构建奠定概念基础。

HCAI 理念的系统化发展经历了从概念提出到框架完善再到范式转变的演进过程。2018 年，斯坦福大学率先建立 HCAI 研究中心，标志着该领域的正式兴起[23]。许为于 2019 年首次提出系统化的 HCAI 框架（图 1a）[4]，其包含三大核心要素：伦理对齐设计，确保 AI 解决方案避免歧视并维护公平正义；技术增强，推动 AI 技术更好地反映人类智能的复杂性和灵活性；人因设计，强调 AI 解决方案的可解释性、可理解性、有用性和可用性。Shneiderman 随后提出了"双维度"设计理念（图 1b）[24]，强调在追求高自动化的同时保持高人类控制，以避免自动化讽刺和过度控制的风险。两个框架均凸显人类在 AI 决策中的最终主导地位，强调 AI 的可解释性、可信赖性与增强作用。目前，全球已有 20 余所高校设立 HCAI 研究机构，形成跨学科研究网络[22,25]。

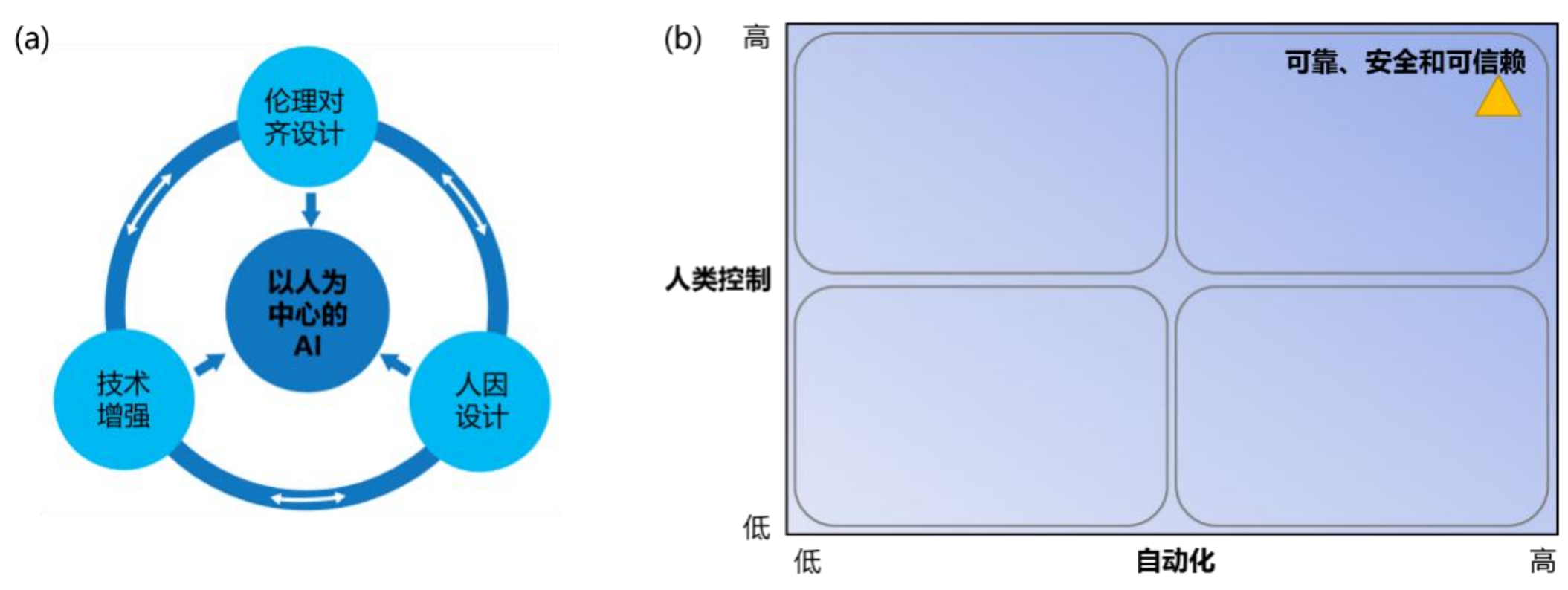

图 1　HCAI 理念发展：（a）许为提出的三要素框架[4]与（b）Shneiderman 提出的双维度框架[24]

为明确 HAII 在 HCAI 体系中的定位，基于智能时代新型人机关系和 HCAI 理念，我们于 2021 年在国内首次系统提出了 HAII 这一跨学科新兴领域[26]。HAII 是指人类与具备自主决策能力的 AI 系统之间的动态交互过程，它超越了传统人机交互中人与被动工具的关系，强调人与 AI 作为认知伙伴的协作模式。与传统非 AI 系统研究相比，HAII 以 AI 系统为核心研究对象，聚焦 AI 所带来的独特挑战与新问题，并探索相应的有效解决路径。其目标是推动 AI 系统与人类的协同研发，优化人-AI 交互，发挥人类智能与机器智能的优势互补，全面纳入伦理考量，并始终坚持人类对 AI 系统的最终控制权。为进一步支撑 HCAI 的跨学科特征，我们于 2024 年提出"人因科学"概念[27]，系统整合工程心理学、人因工程、人机交互、用户体验、神经人因学、宏观工效学等相关领域，从而为 HCAI 研究奠定坚实的跨领域理论和方法论基础。

在此基础上，为确保在人智协作实现以人为中心，我们还提出以人为中心人智协作框架（Human-centered human-AI collaboration framework，图 2）[28]，以"双向赋能"机制阐述了人机协作关系。一方面，人类通过垂直领导模式对 AI 系统进行监督与最终控制，确保关键决策权始终归属人类，尤其是在涉及伦理判断和战略选择时。另一方面，AI 通过变革性领导模式赋能人类，利用其信息处理、学习与计算优势，增强人类的认知能力、决策质量和工作效率[28]。在此基础上，共享责任模式使人类与 AI 能根据各自优势在不同任务阶段动态分配责任，实现优势互补[28]。这种“双向赋能”既确保了人类主导地位和价值取向，又充分发挥 AI 的技术优势，将智能体定位为增强人类能力的伙伴而非替代者，充分体现了以人为中心的 HAII 研究和设计理念[28]。

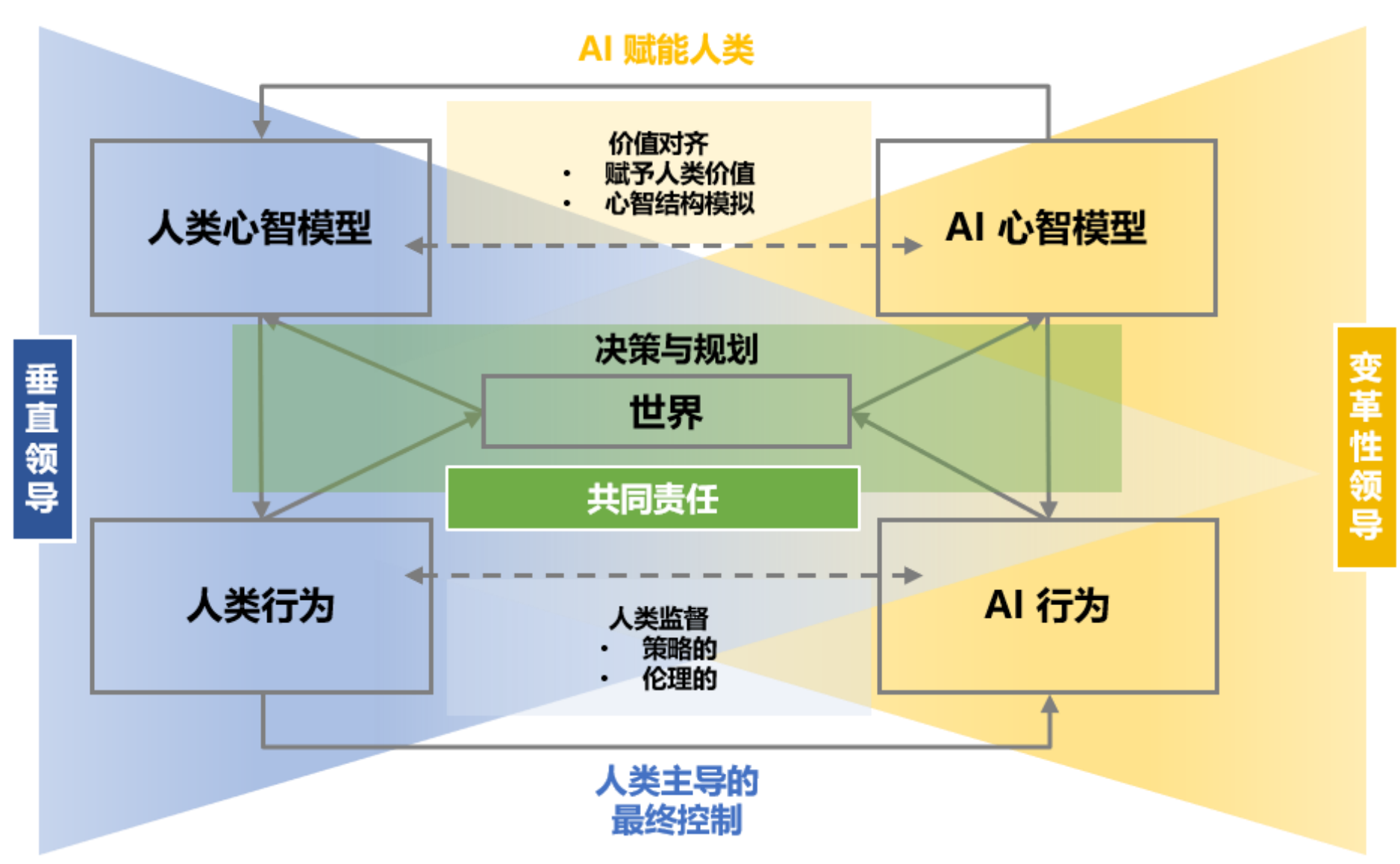

图 2　人类主导的以人为中心人智协作框架[28]。该框架通过双向赋能机制实现真正的人机协作：黄色路径表示 AI 通过价值对齐赋能人类，蓝色路径表示人类对 AI 的最终控制，绿色区域表示动态分配的共享责任

## 2 基础理论研究

在明确研究理念与概念框架后，本节进一步阐述支撑HCAI与HAII的基础理论。人智协同强调，人类与AI并非人机系统中相互独立的两个部分，而是一个整体，其系统绩效取决于复杂动态操作场景下的协作质量。这意味着传统基于信息加工的人机交互视角已难以适应，需要新的理论框架予以支撑。为此，我们引入协同认知系统（joint cognitive system, JCS）的理论视角，并据此提出了三个递进的理论模型，分别对应系统架构、团队协作与社会交互三个维度，为理解和推动高效人智协作提供理论基础。

### 2.1 人智协同认知系统框架

在智能时代，人机关系由传统的单向交互模式转向人智组队（human-AI teaming）的双向协作范式[2–4,8,29,30]。然而，传统人因科学范式在应对这一趋势时存在明显不足：其核心假设是“人–工具”关系，沿用“刺激–反应”的单向交互与静态功能分配，缺乏对 AI 自主性及人–AI 协同机制的系统理解，难以支撑复杂动态环境下的高效、安全协作。

针对这些局限，我们基于协同认知系统理论提出了人智协同认知系统（human-AI joint cognitive system, HJCS）框架[31]。该框架的核心思想是将人–AI 团队视为统一的分析单元，把人类与 AI 视为认知地位对等的合作伙伴，并通过共享认知空间实现深度协同。具体而言，人类操作员与 AI 智能体作为两个认知体（cognitive agents），通过共享态势感知（situation awareness）、相互信任、协同决策与控制实现合作，同时强调人类的主导地位与最终决策权，以确保对 AI 的有效监督与管理。该框架在借鉴 Endsley 态势感知理论的基础上，刻画了人–AI 双方在感知、理解与预测三个层面的信息处理过程，并通过人–AI 合作认知界面实现双向交互。这一模式不仅增强了团队在复杂情境中的态势感知、适应性与执行力，还能根据任务特性和环境变化实现人机角色的智能化调整，从而有效克服传统范式的局限。

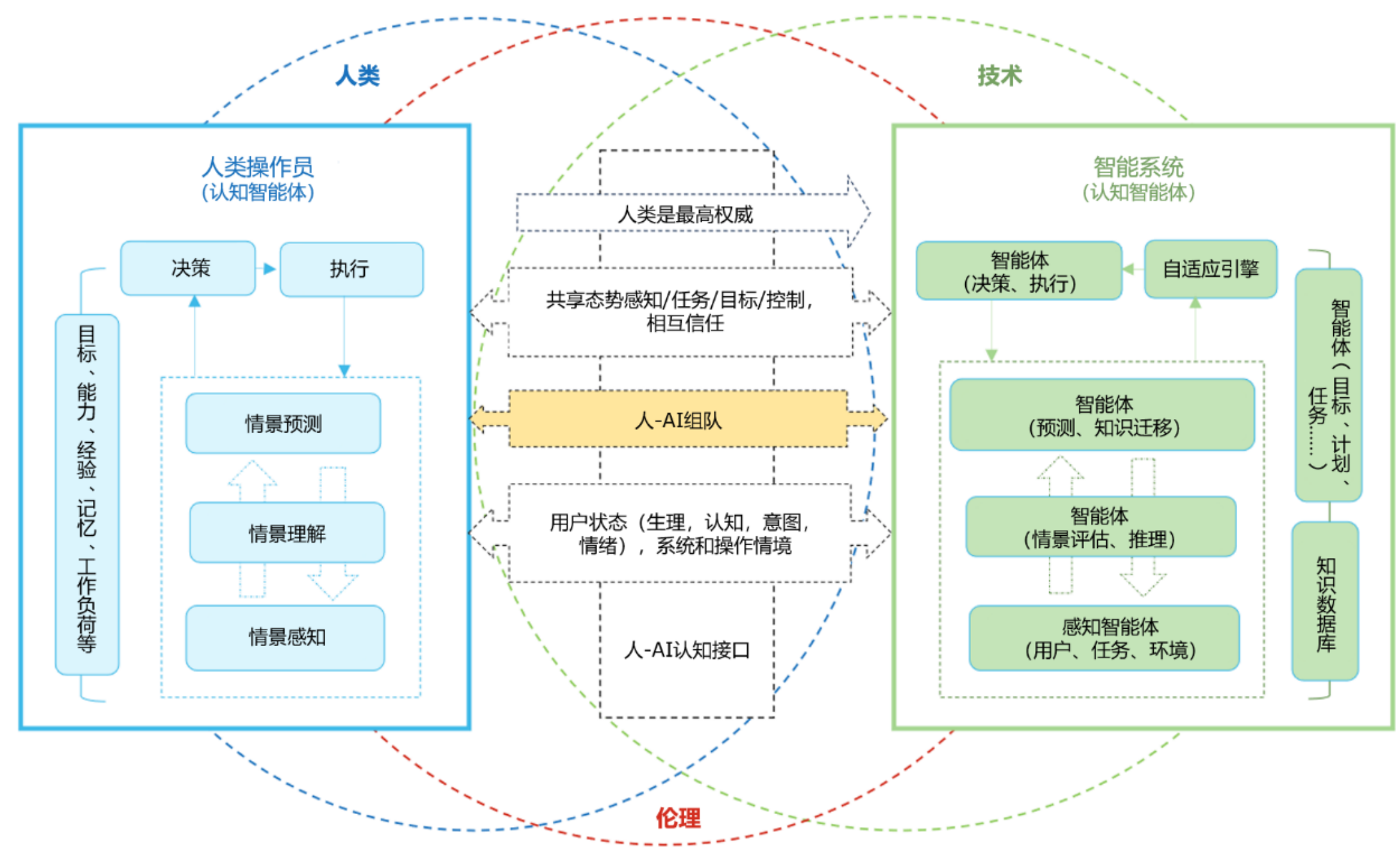

图 3　人智协同认知系统（HJCS）框架[31]。

## 2.2 智能体团队态势感知模型

HJCS 框架确立了人机协同的总体架构，但如何在团队层面实现有效的认知协同仍需进一步理论支持。为此，我们提出了智能体团队态势感知模型（agent teaming situation awareness，ATSA，图 4a）[32]，从认知机制角度深化人-AI 团队协作理论。

ATSA 模型采用双层认知架构：个体层基于 Neisser 的知觉环路理论[33]，描述个体认知主体（人或 AI）如何通过心理模型指导行为，与环境动态交互，形成"认知模型-行动执行-环境感知"的闭环。团队层则描绘多个认知主体协作时如何产生超越个体的集体智能，涵盖团队理解（共享任务和环境认知）、团队控制（协调的行动执行）和外部世界（共同作用的任务环境）三要素。两层间通过交流实现信息传递和认知同步[32]。基于 ATSA 模型的团队控制机制，我们研究了人–AI 协作中控制感的构成与获取机制[34]，发现人–AI 协作中的控制感由四个维度构成：行动自主性（基于个体意愿自主决策与行动的能力）、控制胜任力（对事件与队友实现有效控制的能力）、首要控制策略（在缺乏实际控制时的补偿性策略）与补偿控制策略（通过外部或内部努力实现控制的直接手段）。

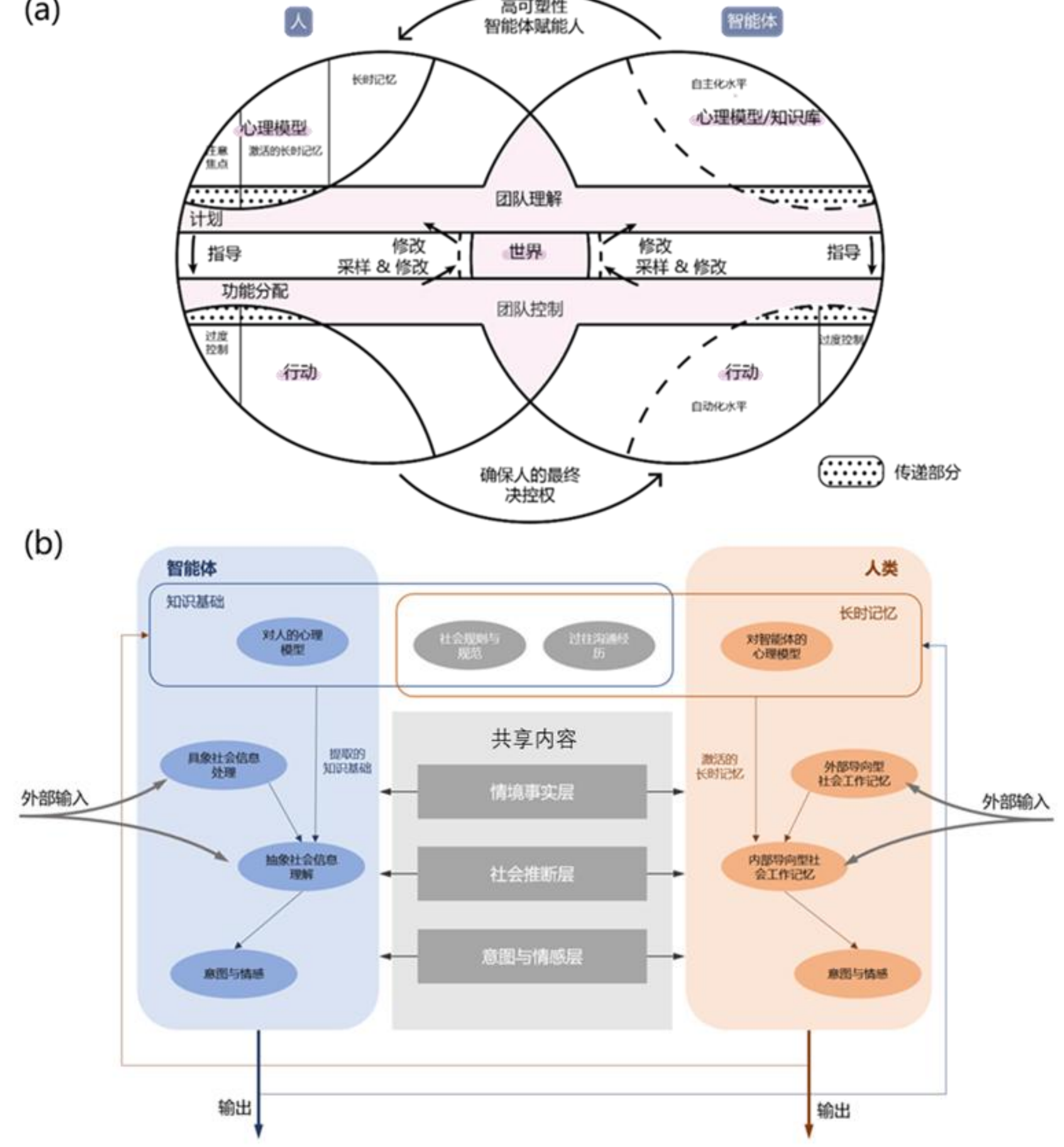

图 4　（a）智能体团队态势感知模型（ATSA）[32]与（b）共享社会理解模型（SSU）。

### 2.3 共享社会理解模型

前述两个模型主要关注功能性协作，但随着 AI 逐渐具备社会性与情感特征，人-AI 交互正从功能性拓展至社会性交互[35–37]。社会性交互依赖复杂的社会线索、文化背景和情感共鸣，对现有 HCAI 理论提出新挑战[38]。为此，我们提出共享社会理解模型（Shared Social Understanding，SSU，图 4b），用于系统描述人–AI 在复杂社会情境下的共享理解建构过程。

为实现有效的人智社会交互，SSU 模型将社会理解过程分解为三个层面，每个层面都要求人机之间达成共享理解：(1)情境事实：侧重对客观、可感知的事实性信息的识别与解析，如识别对话中的关键词、表情、语调等显性信息。(2)社会推断：结合情境特异的长时记忆和社会经验，针对行为动机、价值观念、社会角色、情境约束等深层隐含信息进行推理分析，如理解"微笑"在不同文化背景下的含义差异。(3)意图与情感：在理解前述信息后，人与智能体各自产生内在反应，包括情感体验、行为倾向和未来预期。三个认知过程分别对应人类的外向型与内向型社会工作记忆、意图与情感机制，以及 AI 的具象社会信息处理、抽象社会信息理解与意图-情感生成机制。

上述三个模型共同构建了从宏观架构到微观机制、从功能协作到社会交互的完整理论体系。HJCS 框架提供了顶层系统设计；ATSA 模型揭示了团队态势感知的内在机理；SSU 模型拓展了人–AI 社会性交互的边界。三者相互补充，为 HCAI 与 HAII 的方法论研究、技术开发与应用实践奠定

了理论基础。

## 3 方法论研究

在前述理论基础之上，我们构建了从概念框架到实践路径的方法论体系，以解决 HCAI 理念“如何落地”的问题。尽管 HCAI 理念已逐渐被学界与业界接受，但其具体实施仍缺乏系统化方法论指导。为此，本研究团队提出了涵盖分层实现策略、方法论框架、具体实现方法与组织实践模型的系统化 HCAI 方法论体系。

### 3.1 协同认知生态系统与智能社会技术系统理论

在本次 AI 浪潮中，多智能体系统、群体智能和人-信息-物理系统等新兴技术不断涌现，智能工厂、智能医疗和智能交通等智能生态系统正逐步形成。基于 HJCS 框架，我们提出人-AI 协同认知生态系统（human-AI joint cognitive ecosystems）模型[1,27,39]。该模型将多个 HJCS 视为协同认知网络的组成单元，既保持相对独立性，又通过信息共享、目标协调与资源调配实现生态层级的协同优化，强调跨子系统的相互作用、知识共享与协同进化，从而提升智能生态系统的整体性能。该模型实现了从单一人-AI 协作到生态系统整体协同的递进关系，为 HCAI 和 HAII 提供了更广阔的人因科学支持。

然而，AI 系统的设计、开发与部署涉及技术团队、用户、管理层、政策制定者等多元主体**。**单纯优化技术系统不足以发挥 AI 潜能，必须统筹考虑组织与社会等非技术系统。传统社会技术系统（sociotechnical system, STS）理论强调技术与社会系统的联合优化[40]，但其研究对象主要为非智能系统[41–43]，难以应对 AI 的自主性、学习能力与不确定性[44,45]。因此，我们提出智能社会技术系统（intelligent sociotechnical system, iSTS，图 5）概念[40]。iSTS 在继承传统 STS 理论核心思想的基础上，融入 AI 时代新特性，强调人-AI 协作关系优化、组织再设计、人机共同学习进化、AI 风险管控和开放生态系统中的动态边界管理，从而推动 AI 在宏观社会技术系统环境中的可持续应用。

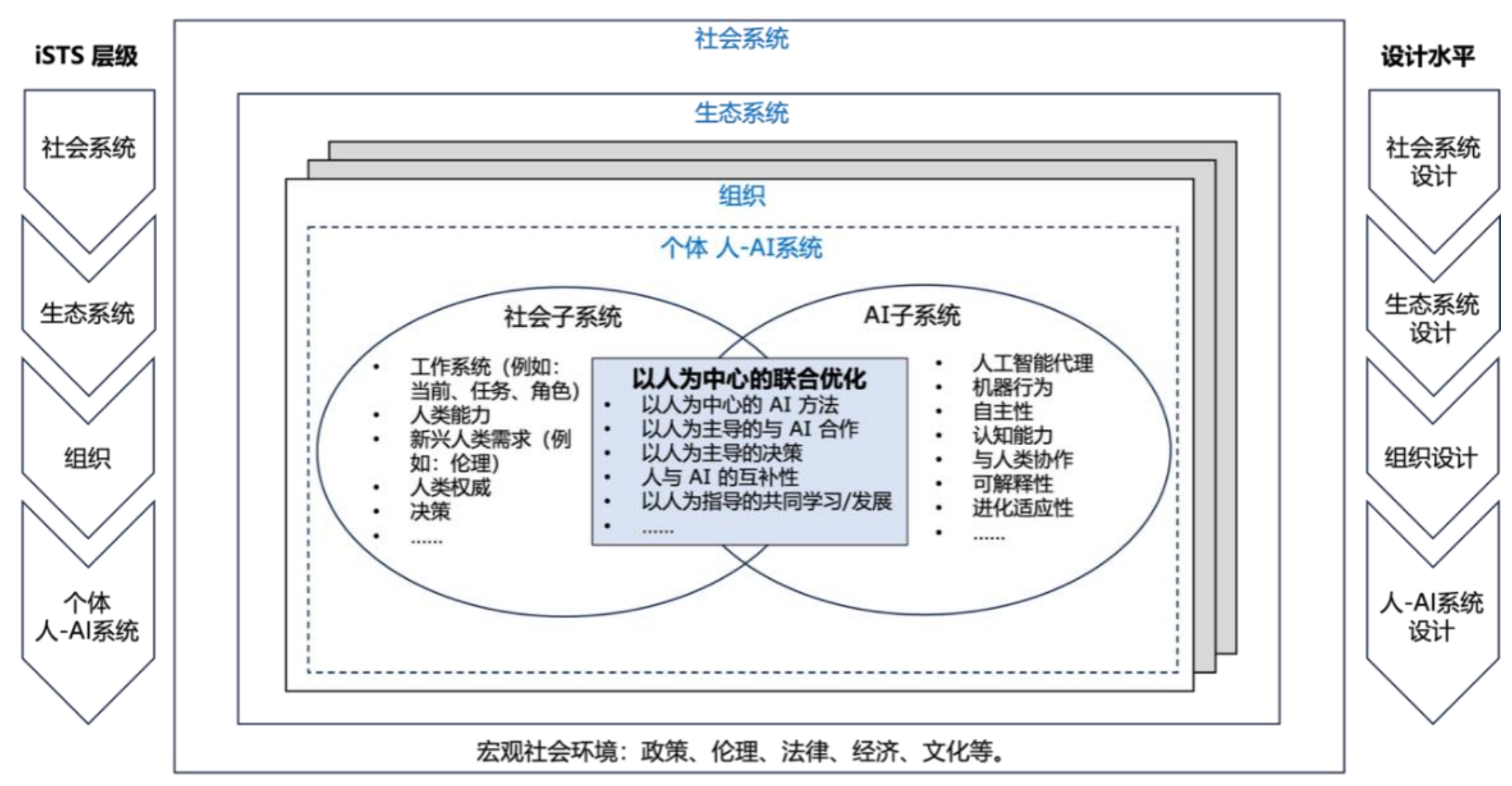

图 5　　智能社会技术系统概念图[40]

### 3.2 层级式 HCAI 框架

在 iSTS 理论指导下，我们进一步提出层级式 HCAI（hierarchical HCAI，hHCAI，图 6）框架[40]。该框架遵循"多层嵌套、逐级扩展"的设计思路，将 HCAI 从个体层扩展至社会系统层，涵盖个体、组织、生态系统和社会四个层次。个体层强调人智协作中的人类最终控制权；组织层推动工作系统再设计（如功能分配、流程优化、技能培训等）；生态系统层促进跨组织与跨系统的伦理对齐；社会层则构建宏观的社会文化、政策和伦理框架。hHCAI 以"组织在环""生态系统在环"与"社会在环"为设计理念，将 AI 开发由传统的"人在环"拓展至整个社会技术环境，确保技术发展与社会价值动态匹配。

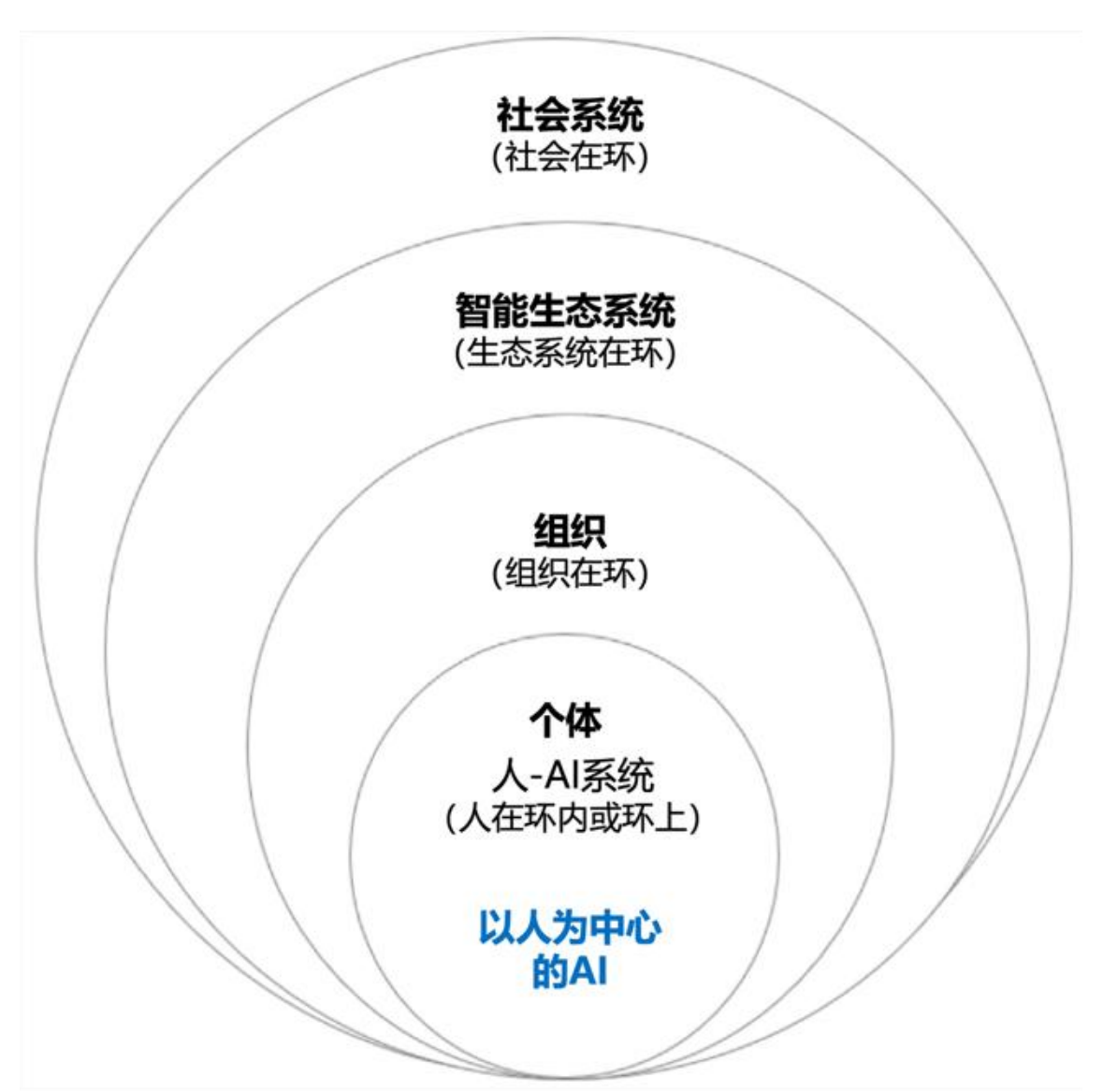

图 6　层级式 HCAI（hHCAI）框架[40]。

**3.3 HCAI 方法论体系**

尽管 hHCAI 提供了宏观指导，但实践仍面临缺乏具体实施路径、协作机制不清晰与质量评估缺失等问题。为解决理论与实践的脱节，我们提出了 HCAI 方法论体系[46]，包括五大核心要素：需求层次、方法分类、过程管理、跨学科协作与多层次设计范式。该体系将战略目标细化为执行路径，融合“双钻式”人本设计流程与 AI 全生命周期流程，并通过单个人智团队、生态系统与社会环境三层策略实现分级落地。

进一步地，我们总结了 HCAI 的五类具体实现方法（图 7）：(1)人本策略类涵盖人类价值对齐、数据和知识双驱动 AI、人机混合增强智能等，确保 AI 系统设计符合人的核心价值和可持续的 AI 发展。（2）计算和建模类包含以人为中心的协同计算和机器学习、可解释 AI 等，强化技术层面对人因需求的响应和协同合作关系。（3）人类可控性类涉及有意义的人类控制、"人在环"设计、基于机器行为管理的可控性设计等，确保系统在关键决策中保持人类主导权。(4)交互设计类重点支持以人为中心的人智协作、人机交互技术和设计、体验设计，促进自然高效的人机协同。(5)标准和治理类包括伦理化 AI 标准、算法与数据治理等，保障 AI 系统的责任性与透明性。该分类体系覆盖 HCAI 的战略规划、技术实现、交互设计及治理监管的完整流程，形成了操作性强的实践指南。

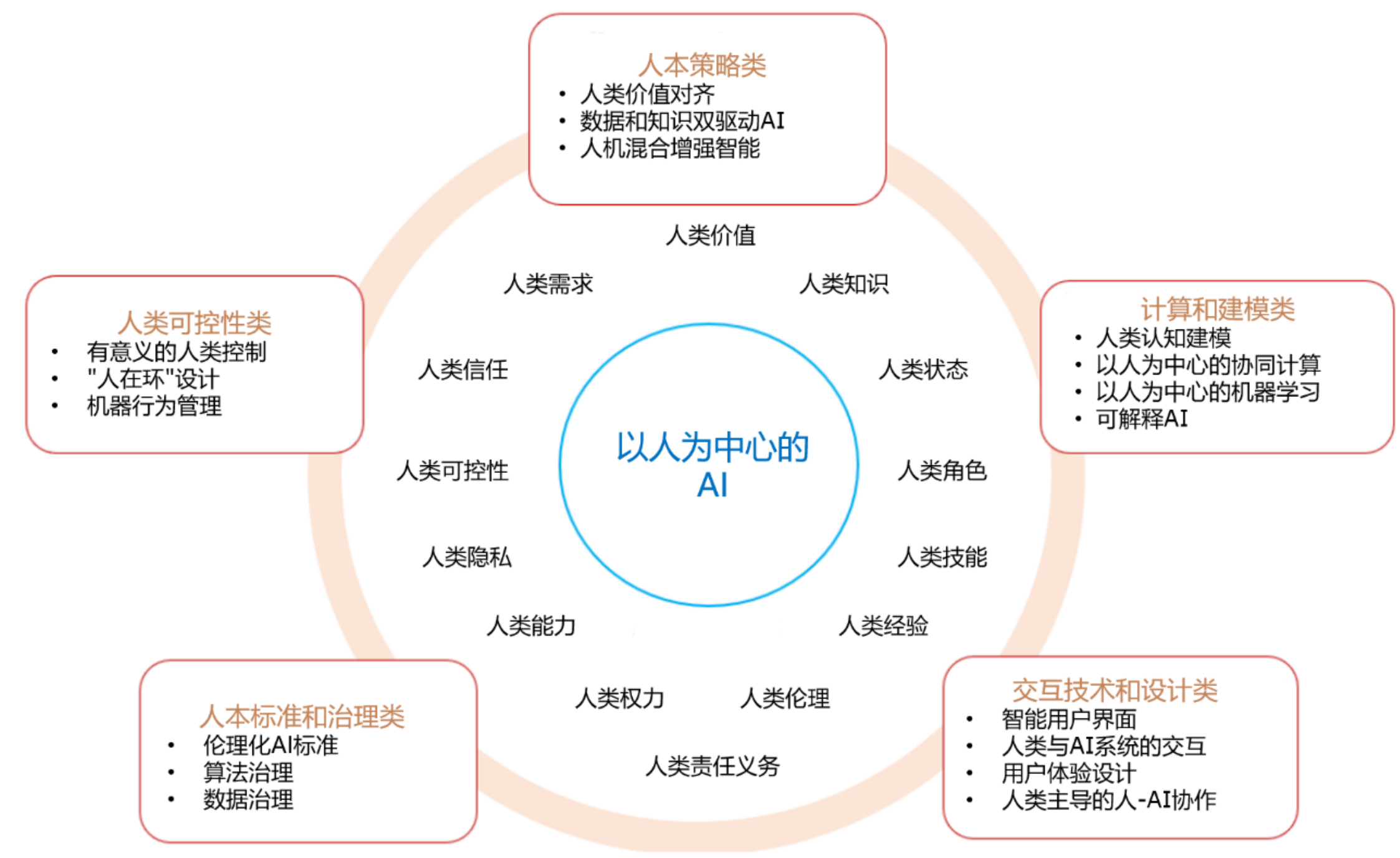

图 7　HCAI 具体实现方法的五大类体系[46]。

## 3.4 组织实践模型

方法论的有效实施最终依赖组织与项目层面的能力支撑。为此，我们与合作团队开发了 HCAI 成熟度模型（HCAI Maturity Model，图 8）[47]。该模型定义了组织 HCAI 能力的五个发展阶段：初始阶段（建立 HCAI 基本认知）、发展阶段（开展试点项目）、定义阶段（建立标准流程）、管理阶段（系统化应用）和优化阶段（持续改进）。每一阶段均有明确的能力要求和评估指标，构建了组织 HCAI 能力从起步到成熟的演进路径，推动 HCAI 由局部试点走向系统整合。此外，考虑到 HCAI 的跨学科特性，我们提出跨学科合作路径，明确了九项具体行动建议，包括构建共享的人本设计理念、应用集成式跨学科方法等，为 HCAI 的可持续发展提供了组织文化与制度保障[48]。

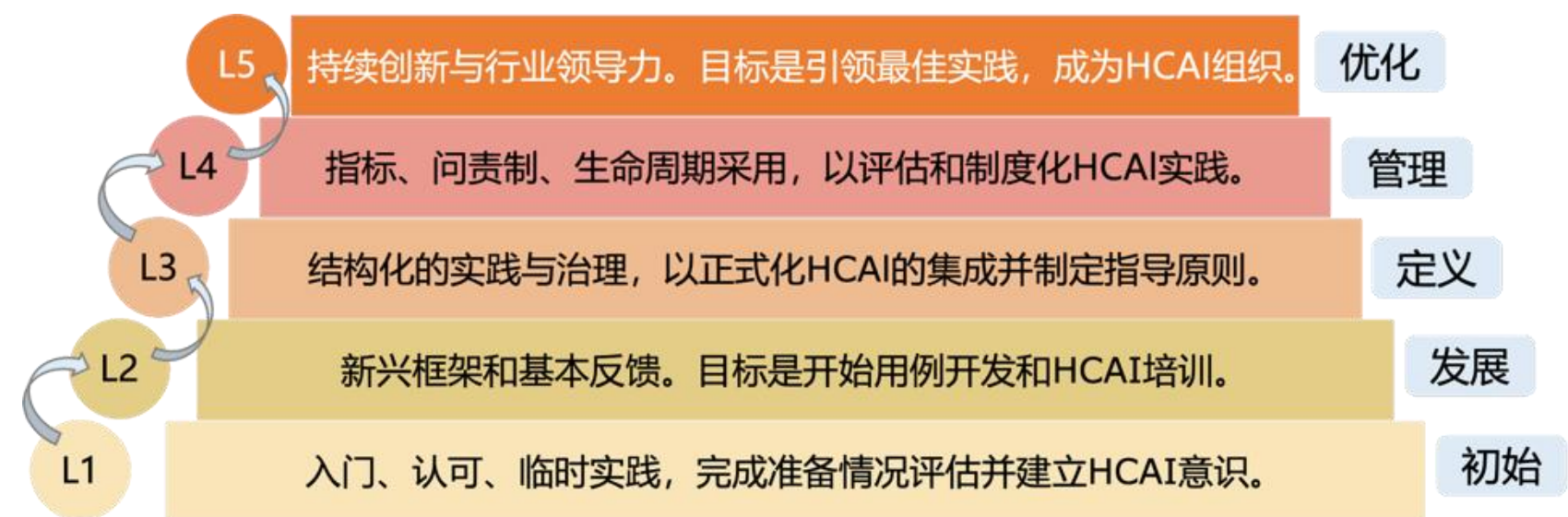

图 8　HCAI 成熟度模型[47]。

# 4 应用实践

在以人为中心理念的引领下，我们将相关理论与方法论逐步应用于多个垂直领域。从自动驾驶中的人车共驾协作，到航空领域的智能驾驶舱设计，再到人机信任机制的构建，这些探索展示了 HAII 理论从抽象框架向具体场景的转化与落地。

## 4.1 面向自动驾驶的三维协同人因策略

自动驾驶作为典型的人-AI 协作系统，是验证 HJCS 框架和 hHCAI 方法论的理想平台。当前 SAE L2-L3 级自动驾驶系统仍需驾驶员有效参与[49]，但现实中存在显著挑战：公众对系统的信任度普遍偏低[50]，频发的安全事故凸显了传统设计的不足[51,52]。现有设计过度强调技术实现，将车载系统视

为工具而非协作伙伴，缺乏 HCAI 理念指导。

为解决这一问题，我们提出了三维协同优化策略[40]：在 HJCS 层面，将车载智能系统定位为认知协作者，基于 ATSA 模型建立双层态势感知机制，实现驾驶员与车辆的双向认知交互（图 9）[53]。在协同认知生态系统层面，构建车–车、车–路与车–云的多主体协作网络，推动从单车智能向网联智能转变。在智能社会技术系统层面，建立技术标准、法规与伦理准则的统一治理框架，确保自动驾驶发展与社会价值观保持一致。该"点-面-体"的渐进实施路径为复杂人-AI 系统的系统性设计提供了可行范式。

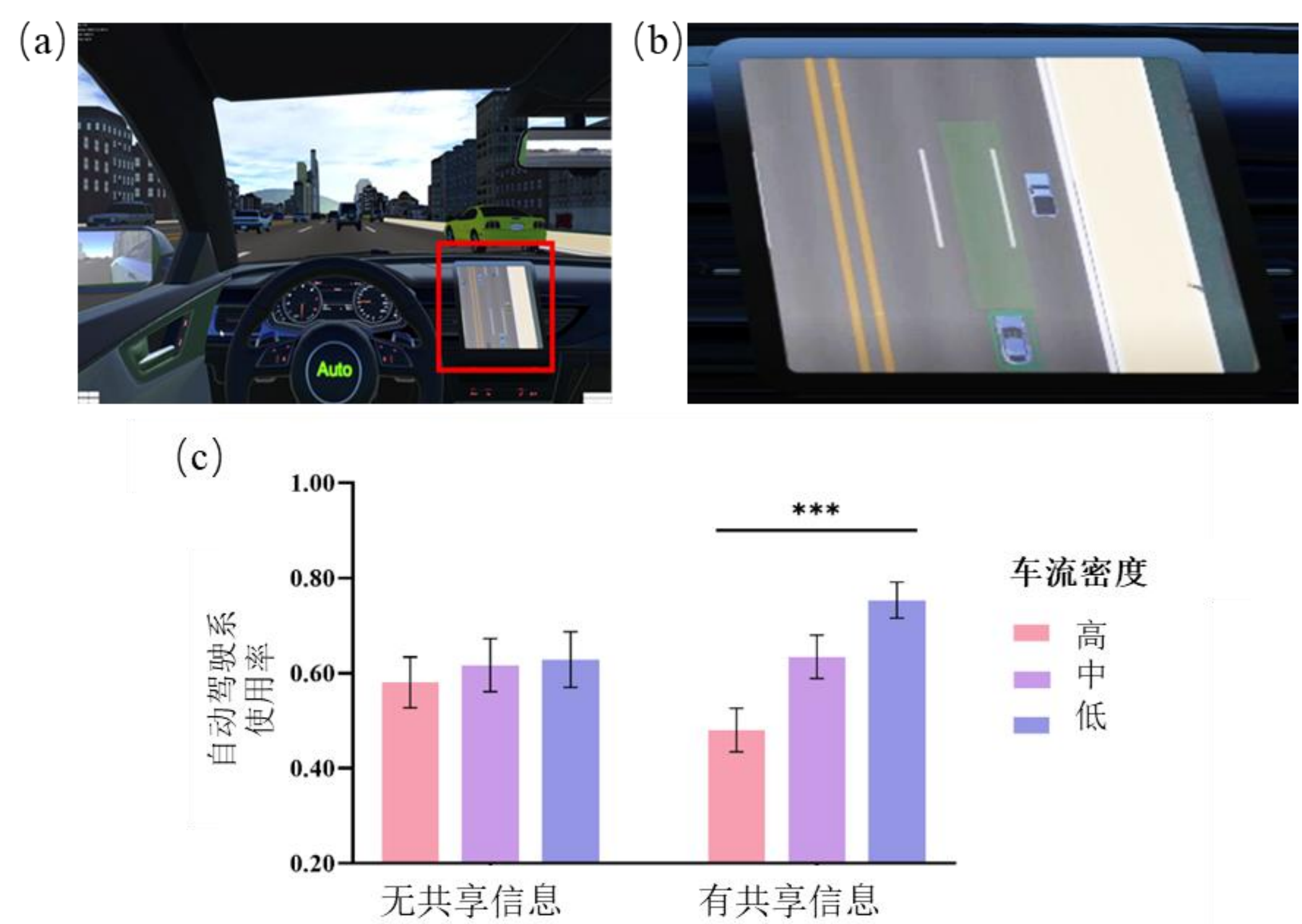

**图 9 基于 HJCS 理念的车载 HMI 设计[53]。（a）将车载智能系统定位为认知协作者，通过 HMI 实时向驾驶员展示 AI 系统对外界信息的感知、理解与预测结果；（b）车载智能系统通过信息共享机制增强人-AI 团队理解；（c）团队理解水平调节驾驶员对自动驾驶系统的信任程度，使驾驶员能够根据车流密度等外界条件动态调整对系统的依赖程度。**

#### 4.2 面向智能飞机驾驶舱的单一飞行员操纵人因研究

航空领域对安全性的极高要求，使其成为检验 HCAI 理论的关键场景。在大型民航智能化驾驶舱单一飞行员操作（Single Pilot Operations, SPO）研究中，我们将以人为中心的理念转化为系统化工程指导原则。在 HJCS 框架和航空安全“零容忍”要求下，我们提出三条设计原则：以飞行员认知边界为约束，构建增强而非替代的协作架构；保障智能座舱系统的可预测性与飞行员最终控制权；建立飞行员–智能系统–地面支持站的三元协同结构，实现信息对称与责任共担[54]。

在系统概念设计中，SPO 设计将机载智能系统重新定义为认知协作者，飞行员与系统通过共享态势感知、协同决策和动态功能分配实现深度合作。从生态系统视角出发，协作范围扩展至地面支持与空中管制，实现跨平台信息共享与协调决策。应用 iSTS 理念，设计同时考虑技术子系统（设备、通信链路）与社会子系统（培训、法规、公众接受度）的联合优化，确保技术进步与社会可接受性匹配。基于此，我们提出了系统透明性设计、动态功能分配、“自动化+自主化”组合方案、人机–空地协同控制与决策、交互优化及飞行员最终控制权保障等具体工程原则。

### 4.3 面向人智协作的信任研究

信任作为人智协作的关键"软"因素，是检验 ATSA 模型和探索人智交互的重要切口。传统研究多关注用户对工具类系统的态度，而在具备自主性与不确定性的 AI 情境下，信任机制更为复杂。我们首次在国内提出人-AI 双向信任问题[54]，并基于 HJCS 和 ATSA 模型开展系统研究[54–56]。

针对传统研究聚焦静态信任结构的不足，我们提出了动态信任发展框架[54]，将信任演化划分为倾向性、初始、实时和事后四阶段，并识别操作者特征、系统特征、情境特征三类关键影响因素，形成"客观特征→主观感知→信任调整"的认知路径。进一步地，我们提出 AI 系统可基于人类状态监测、行为模式分析和任务表现建立对人类的信任模型，实现双向信任机制[54]。在此基础上，我们以自动驾驶为平台开展了信任校准研究（图 9）。我们发现，提升人-智能体间的团队理解可促进信任动态调整[53]。目前，我们正扩展研究至特殊环境（如失重）与特殊人群，以验证信任规律的普适性。

## 5 展望

当前，智能技术发展主要集中在以大语言模型为代表的语言智能领域，并逐步向空间智能（Spatial Intelligence）、生物智能（Biological Intelligence）、具身智能（Embodied Intelligence）等新兴方向拓展。随着人类社会迈向智能体时代，HAII 研究与应用前景广阔，同时也为以人为中心的智能系统发展带来前所未有的挑战与机遇。

### 5.1 亟待突破的科学问题

尽管本文所述理论框架与方法论体系已为 HCAI 与 HAII 研究奠定了坚实基础，但面向未来智能社会仍存在亟需突破的科学难题。

在理论深化层面，现有 HJCS、ATSA 和 SSU 理论框架对动态人机协同的微观认知机制阐释仍需完善。例如，复杂任务流中人类认知状态与 AI 自主行为的实时适配机制尚不清晰，人机控制权动态转移的最优策略、信任多轮演变的精准建模、长期协作中认知负荷的智能分配等核心问题缺乏深入探索。此外，跨文化语境下的 HAII 适用性研究几乎空白，现有理论多基于特定文化背景，对不同社会文化特征下的人机交互范式、伦理价值取向与社会接受度差异缺乏系统性研究。

在方法论完善层面，现有五类 HCAI 实现方法主要针对相对封闭的应用场景，缺乏应对开放环境中突发事件、多目标冲突和伦理困境的动态适应机制。从实验室到真实场景的转化路径仍不明确，纵向数据的不足导致难以系统评估 HAII 系统的长期人因效应，如技能退化、认知依赖、工作满意度变化等潜在风险。更重要的是，跨学科融合仍停留在概念层面，认知科学与 AI 技术的深度耦合尚未实现突破，协同认知系统的计算建模工具链和实验验证平台存在明显断层。

### 5.2 未来发展路径与核心能力建设

面向未来，AI 发展已至关键十字路口。更强大的 AI 系统对监督和控制提出更高要求，任何重大技术决策都可能深刻影响人类未来。因此，以人为中心的 AI 理念愈发重要，如何确保 AI 发展始终服务于人类福祉、有效防范潜在风险，是全球必须共同面对的重大议题。

HCAI 研究需要跨学科深度融合，实践则需要在项目、组织、社会不同层面协同推进。在"大交互"背景下，HAII 研究必须为空间智能、生物智能、具身智能等新兴形态提供跨学科支撑。未来跨学科深度融合的技术路径包括：构建认知-计算-交互一体化的理论建模框架，突破传统学科边界，实现认知科学、计算科学与交互设计的有机融合；开发基于数字孪生的人机协同仿真平台，为复杂 HJCS 系统的设计验证提供高保真测试环境；构建神经符号融合的混合智能架构，结合人类认知的符号推理与 AI 的连接计算优势；形成"认知科学+AI 工程+社会科学"的三螺旋协同创新模式，从组织机制上保障跨学科深度合作。

为此，基于本文提出的理论体系与方法论框架，以人为中心的 HAII 研究需要重点建设五大核心能力体系。在理论模型验证与拓展方面，需通过大规模纵向研究与跨文化对比实验，系统检验和优化 HJCS、ATSA、SSU 等理论框架的适用边界，建立参数化方法，并探索其在新兴智能形态中的扩展应用；在认知计算建模与监测方面，应深化人因科学与 AI 技术的融合，开发基于脑机接口、眼动与生理信号的多模态监测技术，实现对认知负荷、态势感知、信任与意图的精准识别与动态调节；

在跨文化适配能力方面，要构建涵盖价值观、行为模式与交际习惯的跨文化 HAII 知识库，发展文化敏感型设计方法，建立跨文化评估与设计标准；在新型智能可控性方面，须基于 hHCAI 框架研发适配生成式 AI 及具身智能的可解释技术，实现从系统透明到用户理解的跃迁，构建用户可理解、可信任、可控制的交互体验；在全景评估体系方面，应扩展 HCAI 成熟度模型，构建覆盖技术-组织-社会三层的动态评估框架，形成可量化的人本价值指标，并将其嵌入 AI 全生命周期的设计、训练与行为监控环节。

**结语**

以人为中心的 HAII 是智能时代的核心课题之一。其理论体系和实践方法正处于持续建构与迭代中。本文系统梳理了本团队从研究理念到方法论框架，从基础理论到垂直应用的发展脉络，展示了从抽象理念到具体实践的完整转化过程，为该领域的系统性发展提供了参考。未来研究需要在理论深度、方法创新和应用广度三个维度持续推进，以真正实现以人为中心、人机协同共生的智能生态愿景。

**参考文献**